\documentclass[aps,prl,twocolumn,superscriptaddress]{revtex4}

\usepackage[dvips]{graphicx}
\usepackage{amsmath}

\renewcommand{\Vec}[1]{{\bf #1}}
\newcommand{\GVec}[1]{\mbox{\boldmath$#1$}}

\begin{document}


\title{Cyclotron radiation and emission in graphene\\
}



\author{Takahiro Morimoto}
\affiliation{Department of Physics, University of Tokyo, Hongo, 
Tokyo 113-0033, Japan}
\author{Yasuhiro Hatsugai}
\affiliation{Institute of Physics, University of Tsukuba, Tsukuba, 
305-8571, Japan}
\affiliation{Department of Applied Physics, University of Tokyo, Hongo, Tokyo 113-8656, Japan}
\author{Hideo Aoki}
\affiliation{Department of Physics, University of Tokyo, Hongo, 
Tokyo 113-0033, Japan}


\date{\today}

\begin{abstract}
Peculiarity in the cyclotron radiation and emission in graphene 
is theoretically examined in terms of 
the optical conductivity and relaxation rates 
to propose that graphene in magnetic 
fields can be a candidate to realize the Landau level laser, 
proposed decades ago [H. Aoki, Appl. Phys. Lett. {\bf 48}, 559 (1986)].
\end{abstract}


\maketitle


{\it Introduction ---}
There has been an increasing fascination with the physics of 
graphene, a monolayer of graphite, as kicked off by the experimental 
discovery of an anomalous quantum Hall effect(QHE).\cite{Nov05,Zha05} 
The fascination comes from a condensed-matter realization of the 
massless Dirac-particle dispersion at low energy scales on the 
honeycomb lattice,\cite{Mc56,ZhengAndo,Gus05,Nov05,Neto06} 
which is behind all the peculiar properties of graphene.  
In magnetic fields this appears as unusual Landau levels, 
where (i) the Landau levels ($=\sqrt{n}\hbar \omega_c$, $n$: Landau 
index) are unevenly spaced, (ii) the cyclotron frequency 
$\omega_c = (2e/c\hbar)v_F\sqrt{B}$ is proportional to $\sqrt{B}$ 
rather than to $B$, and (iii) there is an extra Landau level right 
at the massless Dirac point ($E=0$), which is outside the 
Onsager's semiclassical quantization.\cite{onsager52}  
While various transport measurements, as exemplified by 
the quantum Hall effect, have been extensively done, optical properties 
are also measured.  For example, Sadowski et al have performed 
a Landau level spectroscopy for a large graphene sample. Inter-Landau 
level transitions are observed at multiple energies, which 
is due to a peculiar optical selection rule ($|n| \leftrightarrow |n|+1$ 
as opposed to the usual $n \leftrightarrow n+1$) as well as 
to the uneven Landau levels.  

Now, if we look at the QHE physics, cyclotron emission from 
the QHE system in {\it non-equilibrium} has been one important phenomenon.  
Experimentally, this typically appears as a strong cyclotron emission 
from the ``hot spot", a singular point in a Hall-bar sample 
where the convergence of electric lines of force puts the electrons 
out of equilibrium.\cite{ikushima}  
Theoretically, one of the present authors proposed a ``Landau-level laser" 
for non-equilibrium QHE systems.\cite{AokiLLL}  
The basic idea is simple enough: we can exploit the unusual 
coalescence of the energy spectrum into 
a series of line spectrum (Landau levels) realize a laser 
from a spontaneous emission if we can make a population 
inversion, where 
the photon energy (= cyclotron energy in this case) is tunable and 
falls on the terahertz 
region for $B\sim 10$T.  However, the most difficult part is 
the population inversion, since if we e.g. optically pump the system, 
the excitation would go up the ladder of equidistant Landau levels 
indefinitely.  

\begin{figure}[tb]
\begin{minipage}{\linewidth}
\includegraphics[width=\linewidth]{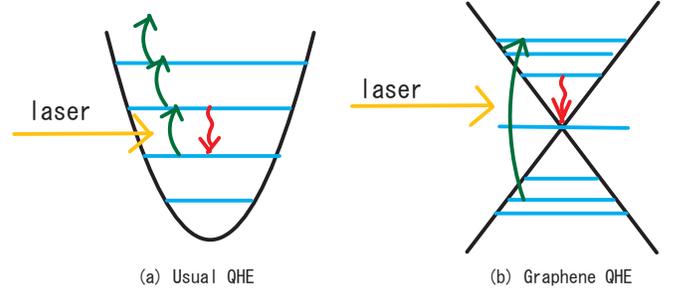}
\end{minipage}
\caption{(Color online) 
Cyclotron absorption (green) and emission (red) processes 
schematically depicted for the ordinary(a) and graphene(b) quantum Hall systems. 
Black lines represent the band dispersion, while blue 
lines Landau levels.
}
\label{radiationscheme}
\end{figure} 

This has motivated us to put a question (Fig.\ref{radiationscheme}): 
will the graphene Landau 
levels, with uneven spacing among their peculiarities, 
favor in realizing such a population inversion?  In this Letter 
we show that this is indeed the case, by actually calculating the optical 
conductivity as well as the relaxation processes.  The message here 
is graphene is a candidate for the Landau-level laser.  

{\it Optical conductivity in graphene ---}
Low-energy physics around the Fermi energy in graphene is described by the 
massless Dirac Hamiltonian,\cite{ZhengAndo}
\begin{equation}
H_0=v_F 
\begin{pmatrix}
0 & \pi^- &0&0 \\
\pi^+ &0&0&0 \\
 0&0&0 & \pi^+ \\
 0&0& \pi^- &0 \\ 
\end{pmatrix} ,
\label{hamiltonian}
\end{equation}
where $v_F$ is the velocity at $E_F$, 
$\pi^{\pm} \equiv \pi_x \pm i\pi_y$, 
$\GVec{\pi}=\Vec{p}+e\Vec{A}$, 
$\Vec{A}$ the vector potential representing 
a uniform magnetic field $\Vec{B}= {\rm rot}\Vec{A}$, 
and the $4\times4$ matrix is spanned 
by the chirality and (K, K') Fermi points.  
In magnetic fields the energy spectrum 
is quantized into Landau levels,
\begin{eqnarray}
\varepsilon_n={\rm sgn}(n)\sqrt{n}\hbar \omega_c, \\ 
\omega_c= \frac{\sqrt{2}}{\ell}v_F = v_F \sqrt{\frac{2eB}{\hbar}},
\end{eqnarray}
for a clean system, 
where $n=0, \pm1, ...$ is the Landau index, and $\ell = \sqrt{\hbar/eB}$ 
the magnetic length.  
Here we consider realistic systems having disorder 
with the self-consistent Born approximation (SCBA) 
introduced by Ando\cite{ando,ZhengAndo} to calculate 
the optical conductivity.

The optical conductivity is given by
\begin{equation}
\begin{split}
\sigma_{\alpha\beta}&(\omega) 
= \frac{e^2\hbar}{i\pi}\int d\varepsilon \frac{f(\varepsilon)}{\hbar\omega}\\
&\times\left[\mbox{Tr}\left( j_{\alpha}{\rm Im}G(\varepsilon)j_{\beta}(G^+(\varepsilon+\hbar \omega)-G^+(\varepsilon))\right) \right. \\
&\hspace{1em} \left. -\mbox{Tr}\left( j_{\alpha}(G^-(\varepsilon)-G^-(\varepsilon-\hbar \omega))j_{\beta}{\rm Im}G(\varepsilon)\right)\right],
\label{kuboformula}
\end{split}
\end{equation}
where $\alpha,\beta = x,y$, $f(\varepsilon)$ the Fermi distribution, and 
$G^{\pm}=G(\epsilon \pm i \delta)$.  For 
Green's function $G$, with the self-energy 
$\Sigma_n(\varepsilon) = 
\Gamma \sum_{n'}[\varepsilon-{\rm sgn}(n)\sqrt{|n|}-\Sigma_{n'}(\varepsilon)]^{-1}$ in the SCBA, the Landau level broadening is given by $\Gamma=n_0V_{0}^2$ 
if we assume for simplicity short-ranged random potential, 
$V=\sum_i V_0\delta(\Vec{r}-\Vec{r}_i)$. 
The light absorption rate is then related to the imaginary part of 
the dielectric function, 
$\varepsilon(\omega)=1+i\sigma_{xx}(\omega)/\varepsilon_0 \omega,$ 
so we can look at $\mbox{Re}\,\sigma_{xx}(\omega)$. 

In order to discuss the optical conductivity in graphene we need 
the current matrix elements across Landau levels.  
The eigenfunctions of the Hamiltonian (\ref{hamiltonian}) 
dictate an unusual selection rule, $|n|-|n'|=\pm1$ 
in place of the ordinary $n-n'=\pm1$, with\cite{ando}
\begin{eqnarray}
j_{x}^{n,n'}=v_F C_{n} C_{n'}
\left[{\rm sgn}(n)\delta_{|n|-1,|n'|}+{\rm sgn}(n')\delta_{|n|+1,|n'|}\right],
\nonumber \\
j_{y} ^{n,n'}=i v_F C_{n} C_{n'}\left[{\rm sgn}(n)\delta_{|n|-1,|n'|}-{\rm sgn}(n')\delta_{|n|+1,|n'|}\right],
\label{matrixelement}
\end{eqnarray}
where
$C_n= 1 (n=0)$ or $1/\sqrt{2}$ (otherwise). 

\begin{figure}[tb]
\begin{minipage}{\linewidth}

\includegraphics[width=\linewidth]{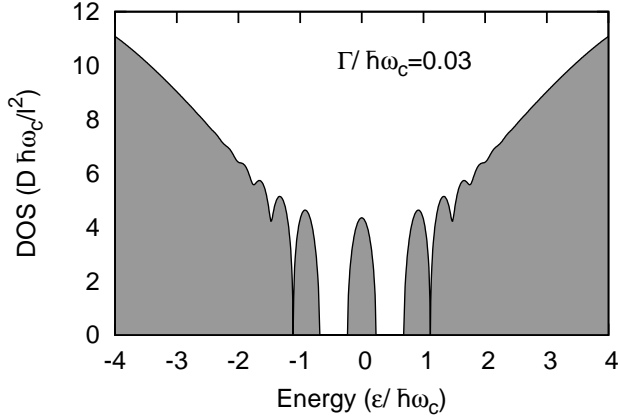}

\end{minipage}
\caption{A typical density of states for graphene 
for an intermediate disorder 
($\Gamma/\omega_c = 0.03$ here). }
\label{dos}
\end{figure}

We have numerically obtained the Green's function and optical conductivity.  
While in usual cases the broadened Landau levels are 
uniformly merged or separated as $\Gamma$ is varied, 
there is a striking difference for graphene, where the Landau levels 
($\propto \sqrt{n}$) are unevenly spaced, so that 
the broadened Landau levels overlap to lesser extent as we go to 
the central one ($n \rightarrow 0$), as typically 
depicted in Fig.\ref{dos}.  Namely, 
for an intermeditate value of $\Gamma/\omega_c$ only the $n=0$ Landau level 
stands alone while the other levels form a continuous spectrum.

\begin{figure}[tb]
\begin{minipage}{\linewidth}

\includegraphics[width=\linewidth]{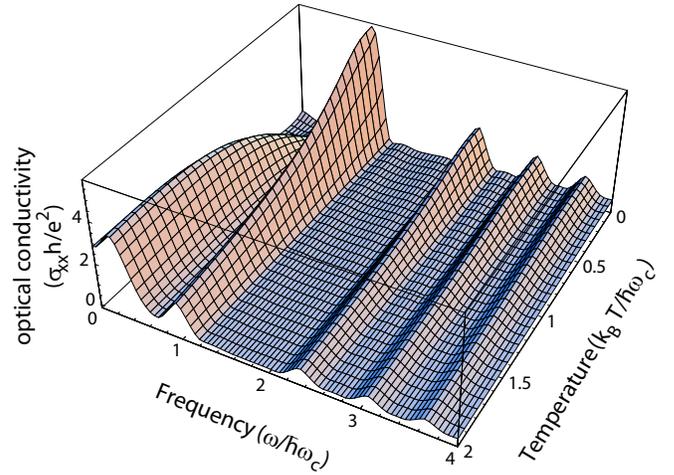}

\end{minipage}

\caption{(Color online) 
Optical conductivity against $\omega$ is shown 
for temperatures $k_BT/\hbar\omega_c = 0 - 2.0$, for a fixed value of 
$\Gamma/\hbar\omega_c=0.01 $ with $\varepsilon_F=0$. 
}
\label{optcon}
\end{figure}

We now look at the optical conductivity in Fig.\ref{optcon} 
for the Fermi energy at 
$\varepsilon_F=0$ (energy for the Dirac point), 
each resonance peak can be assigned to an allowed transition 
with the selection rule (eqn(\ref{matrixelement})).  
The largest peak around $\omega/\omega_c=1$ corresponds to the transition between $n=0 \leftrightarrow \pm1$, while the peaks at higher frequencies 
come from the transition across the Fermi energy, $-n \leftrightarrow n\pm1$.
If we turn to the temperature dependence in the figure, we immediately notice 
a peculiar phenomenon: there is 
a peak, in the region $\omega/\omega_c<1$, 
that grows, rather than decays, for higher $T$.  We can identify this 
as coming from the unusual Landau levels in graphene:  
As $T$ is raised with the Fermi distribution function becoming 
longer-tailed, higher Landau levels begin to be occupied, 
which enables the transitions among higher Landau levels, 
$n \leftrightarrow n\pm1$, to take place.  While this would 
not cause new lines to appear 
for equidistant Landau levels, this does so for the unequally spaced 
Landau levels ($\propto \sqrt{|n|}$) 
for $\omega/\omega_c<1$ transitions.  So we can identify 
this property as one hallmark of the ``massless Dirac" dispersion.

Previously, the optical conductivity has been 
obtained by Gusynin et al,\cite{gusynin}\cite{gusynin-fp} 
who have derived the analytical expression for the optical 
conductivity, 
but the self-energy from the disorder was set to a constant, 
while we have calculated the self-energy self-consistently with SCBA. 
Sadowski et al.\cite{sadowski} also presented a similar expression for the 
conductivity with a constant self-energy as well.  
The present result qualitatively agrees with these, but 
the new findings here are, first, the full dependence 
on the $k_BT/\hbar\omega_c$, including 
the growing of low-frequency peaks at low temperatures.  
Secondly, we point out that the situation as depicted in Fig.\ref{dos} 
should be interesting for 
the cyclotron resonance and emission in non-equilibrium situations 
induced by e.g. an optical pumping with laser beams.  
Namely, the electrons excited to higher energies 
will relax down to the $n=1$ level across the continuum 
spectrum, so that the population inversion across $n=0$ and $n\geq1$ 
should be easier to be realized.  

{\it Relaxation processes ---} 
To quantify this idea, we have to consider the relaxation processes which 
should control the population inversion.  
For the ordinary quantum Hall systems the relaxation processes have been 
extensively discussed.  Specifically, Chaubet et al.\cite{chaubet}\cite{chaubet2} 
discussed dissipation mechanisms, where spontaneous photon radiation and 
coupling with phonons are examined 
on the basis of Fermi's golden rule. 
Other dissipation processes such as electron-electron scattering or impurity scattering, which conserve the total energy, 
do not contribute to inter-Landau level processes 
in the absence of external electric fields (while 
Chaubet et al. have focused on effects of finite electric fields 
in the QHE breakdown where inter-Landau level processes are involved).

So we extend the discussion by Chaubet et al. to relaxation processes 
in graphene.  We first estimate the efficiency of the photon emission with 
Fermi's golden rule: 
$$
W_{i\to f} = 
\frac{2\pi}{\hbar}|\langle i|H_{\rm int}|f\rangle|^2 \delta(\epsilon_f-\epsilon_i).
$$
Here $|i\rangle (\epsilon_i)$ is the wavefunction (energy) in the initial 
state while f stands for the final states, 
and $H_{\rm int}$ the electric dipole interaction 
between the electromagnetic field and electrons.
When the wavelength of light is much larger than the cyclotron radius, 
as is usually the case, we have 
\begin{eqnarray}
W_{i\rightarrow f} &=& \frac{2\pi}{\hbar}\frac V{\pi^2c^3}\int\omega'^2d\omega'
\frac{e^2 \hbar}{2\epsilon_0 V \omega'} 
|\langle i|v|f \rangle |^2 \nonumber \\
&\times& \delta(\hbar\omega'+\epsilon_f-\epsilon_i)
= 4\alpha \left(\frac{|\langle i|v|f \rangle |}{c}\right)^2 \omega, 
\label{emissionrate}
\end{eqnarray}
where $c$ is the velocity of light, $\alpha=e^2/(4\pi\epsilon_0\hbar c)$ the fine-structure constant, 
and we put $\hbar \omega=\epsilon_f-\epsilon_i$ to be 
the cyclotron energy $\hbar \omega_c$.

A peculiarity of graphene appears in the 
current matrix element (eqn(\ref{matrixelement})), for which 
the rate of spontaneous emission, with 
$|\langle n|v|n+1 \rangle | = C_n C_{n+1} v_F$ for graphene plugged in, reads
\begin{equation}
W_{n+1 \to n}^{\rm graphene} =
\begin{cases}
2\alpha \left(\frac{v_F}{c}\right)^2 \omega & (n=0), \\
\alpha \left(\frac{v_F}{c}\right)^2 \omega & (n\neq0).
\end{cases} 
\label{grapheneemissionrate}
\end{equation}
This expression, another key result here, shows that the spontaneous emission rate depends linearly on the cyclotron energy and quadratically on the Fermi 
velocity.  This is in sharp contrast 
with the ordinary QHE systems 
such as the two-dimensional electron gas (2DEG) realized at e.g. 
GaAs/AlGaAs interfaces.  In this case the velocity matrix element 
$|\langle n|v|n+1 \rangle |^2 = (n+1)\hbar \omega /2 m^*$ 
should be plugged in eqn.(\ref{emissionrate}), 
which yields 
\begin{equation}
W_{n+1 \to n}^{\rm GaAs} = 2 (n+1) \alpha \frac{\hbar}{m^* c^2} \omega^2.
\label{GaAsemissionrate}
\end{equation}
This reveals a dramatic difference between graphene and usual 2DEG, 
where the emission rate in the latter is proportional to the square 
of the cyclotron energy.  

We can quantitatively realize the difference: 
The cyclotron energies are
$$
\hbar\omega = 
\begin{cases}
\hbar eB/m^* \sim 1.7 {\rm meV} & {\rm (GaAs)}, \\
v_F\sqrt{2\hbar eB} \simeq 37 {\rm meV} & {\rm (graphene)},
\end{cases} 
$$
for $B = 1$ T, where 
we have adopted the value of graphene Fermi velocity 
$v_F=1.06 \times 10^6$ m/s,\cite{sadowski} 
and the GaAs effective mass $m^* \simeq 0.067 m_e$. 
Hence the cyclotron energy in graphene is orders of magnitude 
larger since it scales as $\sqrt{B}$ reflecting the 
Dirac dispersion, while the energy is usually proportional to $B$.  
If we plug these in eqns.(\ref{grapheneemissionrate},\ref{GaAsemissionrate}), 
we end up with 
$$
W_{i \to f}
\begin{cases}
\propto B^2 \simeq 6 \times 10^4 ({\rm s}^{-1}) & {\rm (GaAs)}, \\
\propto \sqrt{B} \simeq 1 \times 10^7 ({\rm s}^{-1}) & {\rm (graphene)},
\end{cases} 
$$
where the second term in each line indicates the $B$-dependence, 
while the third term a numerical value for $B= 1$ T.  
A conspicuous difference, $\propto B^2$ in the former and 
$\propto \sqrt{B}$, should sharply affect the behavior.  
Thus the spontaneous photon emission rate is orders of 
magnitude enhanced in graphene in moderate magnetic fields 
(as in the above numbers quoted for $B= 1$ T.)  
This indicates that the present system is indeed favorable 
for a realization of the envisaged Landau level laser.

Now, the dissipation process which competes with the photon emission 
is the phonon emission process, which has been discussed 
for the conventional QHE systems, especially 
in the context of the breakdown of the quantum Hall effect\cite{chaubet}. 
The phonon emission rate is also obtained from Fermi's golden rule 
if we replace the electron-light interaction with 
the electron-phonon interaction. 
If we first consider acoustic phonons, the dissipation rate 
is promotional to the extent of the 
overlap between initial and final wavefunctions 
both in usual and graphene QHE systems, which yields a factor 
$e^{-(q\ell)^2}$ 
with $q$ the phonon wavenumber and $\ell$ the magnetic length.  
In usual QHE systems the cyclotron energy 
is $\sim 1$ meV and the magnetic length $\ell=\sqrt{\hbar/eB} \sim 30$ nm 
for $B=$1 T, while the acoustic phonon wavenumber is $\sim 1 $\AA$^{-1}$, 
so that the overlap factor is exponentially small. 
The situation is similar in graphene, 
since the magnetic length $\ell=\sqrt{\hbar/eB}$ is the same.  
So the acoustic phonon emission should be 
negligible in graphene as well in weak electric fields. 
When the applied laser electric field is so intense ($\sim 1$ kV/cm) 
that the Landau levels are distorted and the overlap factor grows, 
the phonon emission may begin to compete with the photoemission.

Are there any other factors that distinguish graphene from 2DEG's?  
In this context we can note that 
Chaubet {\it et al.} have further pointed out the following.  
In an electron system 
confined to 2D a wavefunction has a finite tail in the 
direction normal to the plane, and 
the phonon emission is enhanced through the coupling of 
the tail of the wavefunction and perpendicular phonon modes 
which propagate normal to the 2D system in 
the substrate\cite{chaubet2}. This way 
the phonon emission can compete with the spontaneous emission in usual 
QHE systems.  
By contrast, a graphene sheet is an atomic monolayer, and 
there is only a loose coupling with the substrate.  
We can also consider acoustic phonons coupled with impurity scattering, 
which may compensate the momentum transfer $q$ of phonons, and hence 
the overlap factor $e^{-(q\ell)^2}$.\cite{comment2}
To be precise, graphene itself should have phonon modes that include 
the out of plane modes, and their effects is an interesting 
future problem.
As for optical phonons, their energies are known to be 
higher than 100 meV for wavelength $q=0$ in graphene\cite{opticalphonon}, 
so that optical phonons do not contribute to the dissipation 
for $B \sim$ a few tesla with $\hbar\omega\sim$ 40 meV.  
Overall, we conclude that the dissipation due to acoustic phonons will be small 
in graphene in the weak electric-field regime.  
When the pumping laser intensity is not too strong to invalidate 
the present treatment but strong enough for the population inversion, 
the present reasoning should apply, and we can expect 
efficient cyclotron emissions from graphene.  

Entirely different, but interesting is 
the problem of Anderson localization arising from disorder.  
While this is out of scope of the present work, 
we can expect delocalized states with 
diverging localization length at the 
center of each Landau level are present as inferred from the QHE observation, 
whose detail is an interesting future problem.  
The situation should also depend on whether the disorder 
is short-range or long-range, but, in ordinary QHE systems, 
a sum rule guarantees the 
total intensity of the cyclotron resonance intact.\cite{AokiLLL}  
As for the ``ripples", suggested to exist in actual graphene 
samples\cite{meyer2007}, the $n=0$ Landau level remains sharp 
(which is topologically protected since the slowly varying 
potential does not destroy the chiral symmetry\cite{hatsugai_fukui_aoki}), 
while other levels become broadened,\cite{giesbers2007} 
and this favors the situation proposed in the presented paper.\cite{comment1}

{\it Summary ---} 
To summarize, we have discussed 
the radiation from graphene QHE system.  
We conclude that unusual uneven Landau levels, 
unusual cyclotron energy, unusual transition selection rules 
all work favorably for a  population inversion envisaged 
for the Landau level laser.  
An estimate of the photon emission rate shows that 
the emission rate is orders of magnitude more efficient than in 
the ordinary QHE system, while the competing phonon emission 
rate is not too large to mar the photon emission.  
Important future problems include the examination of the actual 
lasing processes including the cavity properties, coupling of 
electrons to the out-of-plane phonon modes, etc.  
We wish to thank Andre Geim for illuminating discussions.
This work has been supported in part by Grants-in-Aid for Scientific 
Research on Priority Areas from MEXT, ``Physics of new quantum phases in 
superclean materials" (Grant No.18043007) for YH, 
``Anomalous quantum materials" (No.16076203) for HA.


\begin{thebibliography}{20}
\bibitem{comment2} 
Localized states around point defects were detected recently with 
STM in $B$ \cite{niimi}, 
and their radius was found to be comparable with the 
magnetic length ($\sim 30$ nm).
\bibitem{comment1} 
We can also note that 
the $n=0$ Landau level in graphene has an exactly $E=0$ 
edge mode, of a topological origin, right at the 
center\cite{hatsugai_fukui_aoki}, which may also affect the 
photon absorption/emission.
\end{thebibliography}
\end{document}